\documentclass[prd,floats,floatfix, twocolumn,
superscriptaddress,nofootinbib,showpacs]{revtex4-1}

\usepackage[english]{babel}

\usepackage[letterpaper,top=2cm,bottom=2cm,left=3cm,right=3cm,marginparwidth=1.75cm]{geometry}

\usepackage{amsfonts}
\usepackage{amsmath}
\usepackage{amssymb}
\usepackage{amsthm}
\usepackage{bm}
\usepackage{dcolumn}
\usepackage{epsfig}
\usepackage{graphicx}
\usepackage{graphics}
\usepackage{latexsym}
\usepackage{rotating}
\usepackage{comment}
\usepackage{xspace}
\usepackage{soul}

\usepackage[dvipsnames, usenames]{xcolor}
\definecolor{linkcolor}{rgb}{0.0,0.3,0.5}
\usepackage[hypertexnames=false, unicode, colorlinks=true, linkcolor=linkcolor,
citecolor=linkcolor, filecolor=linkcolor,urlcolor=linkcolor,
pdfusetitle]{hyperref} 

\usepackage{mathrsfs}
\newcommand\ml{\mathcal}
\newcommand\nn{\nonumber}
\newcommand\bea{\begin{eqnarray}}
\newcommand\eea{\end{eqnarray}}

\begin{document}

\title{An observational signature for extremal black holes}
\newcommand{\URI}{\affiliation{Department of Physics, 
    University of Rhode Island, Kingston, RI 02881, USA}}    
\newcommand{\URICCR}{\affiliation{Center for Computational Research, 
    University of Rhode Island, Kingston, RI 02881, USA}}    
\newcommand{\UMassDPhy}{\affiliation{Department of Physics, 
    University of Massachusetts Dartmouth, 285 Old Westport Rd., North Dartmouth, MA 02747, USA}}    
\newcommand{\UMassDCSCDR}{\affiliation{Center for Scientific Computing \& Data-Science Research, 
    University of Massachusetts Dartmouth, 285 Old Westport Rd., North Dartmouth, MA 02747, USA}}    

\author{Stefanos Aretakis\thanks {aretakis@math.toronto.edu}}\affiliation{\small Department of Mathematics, University of Toronto, 40 St George Street, Toronto, ON, Canada}
\author{Gaurav Khanna \thanks {gkhanna@uri.edu}}
\URI
\UMassDPhy
\UMassDCSCDR
\URICCR
\author{Subir Sabharwal \thanks {subir@uri.edu}}
\URICCR

\begin{abstract}
We consider scalar perturbations of the Reissner--Nordstr\"{o}m family and the Kerr family. We derive a characteristic expression of the radiation field, \textit{at any given fixed angle} of future null infinity, and numerically show that its amplitude gets excited only in the extremal case. 	Our work, therefore, identifies an observational signature for extremal black holes. Moreover, we show that the source of the excitation is the extremal horizon instability and its magnitude is exactly equal to the conserved horizon charge.
\end{abstract}

\maketitle

\section{Introduction}
\label{intro}
Black holes are one of the most interesting and important predictions of general relativity. In this paper, we study signals that could potentially be received by measurements on Earth leading to the detection of black holes and further understanding of their properties. Specifically, we address the problem of deriving \textit{observational signatures} of black holes, namely the problem of finding distinguishing features of \textit{the dynamics of perturbations of black holes} which can be measured and studied by far-away observers. 

In this paper, we focus on observational signatures of a special class of black holes, the so-called \textit{extremal black holes} -- specifically extremal Reissner--Nordstr\"{o}m (ERN) and extremal Kerr (EK). These black holes play a fundamental role in mathematical relativity, astrophysics and high energy physics. From a theoretical point of view, they saturate geometric inequalities for the mass, angular momentum and charge~\cite{dainprl6,alaeeprl}. Furthermore, they play an important role in the study of Hawking radiation and more generally in quantum gravity \cite{haw95}, and in string theory \cite{stromingerextremalentropy} due to the fact that they have zero temperature. Their near-horizon geometry produces new solutions to the Einstein equations with conformally invariant properties \cite{nearhorizonrn}. Both gravitational and electromagnetic signatures of the near-horizon geometry have been introduced in \cite{gralla2016, grallastrominger}. On the other hand, from a more practical point of view, astronomical evidence suggests that near-extremal black holes are ubiquitous in the universe. Evidence supports the existence of both stellar near-extremal black holes \cite{rees2005} and super-massive near-extremal black holes \cite{brennemanpaper}. Even though, the spins of the astrophysical black holes in all the aforementioned works are below the upper bound called the Thorne limit \cite{Thorne1974}, recent works open the possibility that there are astrophysical black holes beyond the Thorne limit \cite{Skadowski2011}. \textit{The present work provides an additional tool in the quest of finding black holes that are near or beyond the Thorne limit}. 

Perturbations of extremal black holes exhibit a \textit{horizon instability} which manifests itself in the growth along the event horizon of derivatives transversal to the event horizon \cite{aretakis1, aretakis2, aretakis4, hj2012}. This dynamical behavior of extremal black holes originates from a characteristic conserved charge along the event horizon, known as the horizon charge $H[\psi]$, where $\psi$ is a scalar perturbation. The existence of this charge calls into question \textit{the no-hair hypothesis}~\cite{nohair,nohair2,nohair3} for extremal black holes according to which the only externally observable classical parameters of a black hole spacetime are the mass, angular momentum and the electric charge. On the other hand, we note that it has been shown \cite{aretakis1} that all translation-invariant quantities decay in time away from the extremal horizon; hence, the conserved charge $H[\psi]$ leaves a non-decaying trace of the perturbation \textbf{only} along the horizon and for this reason it can be regarded as \textit{horizon hair}. In other words, \textit{the charge $H[\psi]$ could be viewed as yet another classical parameter of the extremal black hole spacetime}. We remark that the observational signature of the present paper relies precisely on the existence of $H[\psi]$. This work provides numerical evidence for the measurability of the horizon hair $H[\psi]$ by external observers, including far-away observers, potentially falsifying the no-hair hypothesis in the case of extremal black holes. {\color{black}Note, however, that our work is in the context of an idealized isolated black hole in the framework of black hole perturbation theory i.e. a fixed background black hole spacetime. It does not apply to the case of full non-linear Einstein field equations of general relativity.}

The proposed signature is observable by a single detector at null infinity i.e. a fixed angle; as opposed to a previously published expression that requires an integral over the entire $S^2$ sphere~\cite{aretakis5}. {\color{black} This is significant, since it brings the in-principle observation of the hair $H[\psi]$ closer to what would be possible to do on physically realistic grounds.}

{\color{black} The approach we take in this paper is somewhat atypical of the approach taken in the related mathematical relativity literature. We propose a conjecture which we do not attempt to prove rigorously; rather we provide compelling numerical evidence in support of it. In particular, the conjecture presented is an explicit expression for a signature $s_{\mathcal{I}^{+}}$ based on the scalar radiation field along a fixed null generator of $\mathcal{I}^{+}$. {\em This signature is verified to be numerically zero for subextremal black holes (RN and Kerr), and shown to equal $H[\psi]$ in the extremal cases}. This is the main result presented in this work. Additionally, it should be noted that while the key expressions and discussion presented is done in the context of RN black holes, the major result is also generalized to the Kerr case towards the end of the {\em Numerical Results} section.}

Finally, it is worth noting that the same notion of horizon hair $H[\psi]$ approximately applies in the case of near extremal black holes, as a {\em transient} with duration dependent on the proximity to extremality. Therefore, we expect the results of this work to similarly hold in the near extremal cases, especially for cases wherein the charge or spin is within $5\times10^{-5}M$ of the extremal value~\cite{subir2}. We also remark that, under the presence of superradiance, an amplified version \cite{dejan2023, zimmerman1} of the horizon instability on such backgrounds due to the existence a sequence of zero-damped quasinormal modes is already known \cite{glampedakisfull,zeni13}.

\section{The geometric setting (RN \& Kerr)}
\label{sec:TheGeometricSetting}
Our goal is the measurement of the horizon hair $H[\psi]$  by far-away observers who receive radiation from the near-horizon region. In our model, observers are assumed to live on null generators of the future null infinity $\mathcal{I}^{+}$ of either the Reissner--Nordstr\"{o}m family or the Kerr family. The future null infinity $\mathcal{I}^{+}$ admits the following coordinates
\[\mathcal{I}^{+}=(\tau, \vartheta), \]
where the coordinates are defined as follows
\begin{itemize}
	\item $\tau$: the \textit{retarded time} (on $\mathcal{I}^{+}$) or \textit{advanced time} (on $\mathcal{H}^{+}$),
	\item $\vartheta=(\theta^1,\theta^2)$: the \textit{angle} which represents the direction of any given ideal observer along $\mathcal{I}^{+}$. 
\end{itemize}
Null generators of $\mathcal{I}^{+}$ are parametrized by 
\[ \{(\tau, \vartheta_{0}), \tau\in\mathbb{R} \}\]
where $\vartheta_{0}$ is a fixed angle.

{\color{black}
\section{Summary of previous work (RN)}
In this section, we briefly summarize the rigorously derived results for leading and sub-leading tail behavior for scalar fields at null infinity, $\ml{I}^{+}$ for sub-extremal and extremal RN black holes. In~\cite{logasymptotics}, the leading and sub-leading behavior of scalar fields was derived for sub-extremal RN BH's. Specifically, this was done for four different types of initial data (ID). In this work, we will focus only on the astrophysically relevant ``{\bf{Type A}} ID'' (see~\cite{paper4} for definitions of ID) where the Type A ID is (i) horizon ($\ml{H}^+$) penetrating, (ii) with compact support and (iii) no support on $\ml{I}^+$.\\

We consider massless scalar perturbations $\psi$ on either the Reissner--Nordstr\"{o}m (RN) family or the Kerr family:
\begin{equation}
\Box_g \psi=0
\label{wave}
\end{equation}
where $g$ is the RN/Kerr metric. 
Following Ref.~\cite{logasymptotics}, we define the radiation field $\Psi$ of the scalar field $\psi$ as follows,
\[\Psi(\tau,\vartheta)=\lim_{r\rightarrow \infty}(r\psi)(\tau, r,\vartheta).\]
For {\bf{Type A}} ID, the authors derive in~\cite{logasymptotics} that the asymptotic tail behavior for sub-extremal RN is given by
\bea
\Psi(\tau,\vartheta)=  C_1(\vartheta)\frac{1}{\tau^2}+C_2(\vartheta)\frac{\log(\tau/M)}{\tau^3}+O\left(\frac{1}{\tau^3}\right)\nn
\eea
They further proved that for generic non-spherically symmetric perturbations, the coefficients $C_1({\vartheta})$ and $C_2({\vartheta})$ {\it do not} depend on $\vartheta$ and only the spherically symmetric $\ell=0$ field component contributes to the value of these coefficients. Indeed the angular independence is consistent with Price's law that states that for spherically symmetric BH space-times like RN, the leading order tail behavior at $\ml{I}^+$ for a generic perturbation with $\ell\geq1$ is given by $1/\tau^{(\ell+2)}$ \cite{leaver}.\\

The above results are extended to ERN in Ref.~\cite{paper4}, specifically deriving late-time asymptotics for the various ID. But, for {\bf{Type A}} ID, the authors were only able to derive the leading order behavior of $C_1(\vartheta)/\tau^2$. Additionally, they were able to prove that for the generic {\bf{Type A}} perturbation, only the spherically symmetric component, $\ell=0$, would contribute to the coefficient $C_1(\vartheta)$. For {\bf{Type B}} ID, which is compactly supported away from $\ml{H}^+$ and not supported on $\ml{I}^+$, they were able to prove that the sub-leading contribution is of the logarithmic form, $\log(\tau/M)/\tau^3$ and provide its coefficient explicitly. Again, the coefficient $C_2(\vartheta)$ is independent of $\vartheta$ on $\ml{I}^+$ and consistent with Price's Law since $\ell\geq1$ would contribute a term decaying as $1/\tau^3$ or faster. \\

Given the similarities between {\bf{Type A/B}} ID, including in the vanishing of the Newman-Penrose constant $I_0[\psi]$ -- {\it we conjecture the existence of the $\log(\tau/M)/\tau^3$ term as the sub-leading contribution to the scalar tail behavior for both ERN and EK and further that $C_2(\vartheta)$, the coefficient of the $\log$ term, is dependent only on the spherically-symmetric $\ell=0$ piece of the initial perturbation motivated by Price's Law.} These taken together, prompt a proposed expression for $C_2(\vartheta)$ that is presented in the next section. The main contribution of this work then is to numerically verify the conjectured expression for $C_2(\vartheta)$ for ERN and the existence of a similar expression for EK BH's. Finally, the existence and form of the $C_2(\vartheta)$ term allows us to posit an {\it anisotropic observational signature at $\ml{I}^+$ for the extremality of BHs}. 
}

\section{Observational signatures in bounded regions (RN)}
\label{hairhhbar}
Observational signatures can easily be derived in bounded regions from known results. It should be noted that unless explicitly stated otherwise, the expressions presented in this section are for RN. Starting first from the event horizon, recall \cite{paper4} that asymptotically as $\tau\rightarrow\infty$ we have
\begin{equation*}
   \psi|_{\mathcal{H}^{+}}(\tau,\vartheta)\sim \begin{cases}
8I^{(1)}[\psi]\cdot \frac{1}{\tau^3} & \text{on subextremal RN}\\
\frac{2}{M}H[\psi]\cdot \frac{1}{\tau} & \text{on ERN}
    \end{cases}
\end{equation*}
where  $H[\psi]$ is defined below,  
\begin{equation}
H[\psi]:=-\frac{M^2}{4\pi}\int_{\rm EH}\,\partial_r(r\psi)\,d\Omega\, 
\end{equation}
and $I^{(1)}[\psi]$ is defined in the following section. Defining the signature $s_{\mathcal{H}^{+}}(\vartheta)$ as follows 
\begin{equation}
s_{\mathcal{H}^{+}}(\vartheta)=\lim_{\tau\rightarrow \infty} \frac{M}{2}\tau\cdot \psi|_{\mathcal{H}^{+}}(\tau,\vartheta)
\label{s2ern}
\end{equation}
then 
\begin{equation*}
   s_{\mathcal{H}^{+}}(\vartheta)= \begin{cases}
0 & \text{for subextremal RN}\\
H[\psi] & \text{for ERN}
    \end{cases}
\end{equation*}
We next consider the constant $r$ hypersurface $R_{r_0}=\{r=r_0\}$. From \cite{paper4} it follows that asymptotically along $R_{r_0}$ we have 
\begin{equation*}
   \psi|_{\{r=r_0\}}(\tau,\vartheta)\sim \begin{cases}
8I^{(1)}[\psi]\cdot \frac{1}{\tau^3} & \text{on subextremal RN}\\
\frac{4MH[\psi]}{r_0-M}\cdot \frac{1}{\tau^2} & \text{on ERN}
    \end{cases}
\end{equation*}
Defining the signature $s_{r_0}(\vartheta)$ as follows 
\begin{equation}
s_{r_0}(\vartheta)=\lim_{\tau\rightarrow \infty} \frac{r_0-M}{4M}\tau^2\cdot \psi|_{\{r=r_0\}}(\tau,\vartheta)
\label{s2ern}
\end{equation}
then 
\begin{equation*}
   s_{r_0}(\vartheta)= \begin{cases}
0 & \text{for subextremal RN}\\
H[\psi] & \text{for ERN}
    \end{cases}
\end{equation*}

\section{The observational signature $s_{\mathcal{I}^{+}}(\vartheta)$ at null infinity (RN)}
It should be noted that unless explicitly stated otherwise, the expressions presented in this section are for RN. From the radiation field $\Psi(\tau,\vartheta)$, we extract the following limits (assuming they exist)
\begin{equation}
C_1(\vartheta)= \lim_{\tau\rightarrow \infty}\tau^2\cdot\Psi(\tau,\vartheta)
    \label{C1def}
\end{equation}
and
\begin{equation}
C_2(\vartheta)=\lim_{\tau\rightarrow \infty}\frac{\tau}{\log(\tau/M)}\cdot \left(\tau^2\cdot\Psi(\tau,\vartheta)-C_1(\vartheta)\right).
    \label{C2def}
\end{equation}
Clearly, both $C_1(\vartheta), C_2(\vartheta)$ can be computed by knowing the radiation field $\Psi(\tau,\vartheta)$ along a fixed null generator of null infinity corresponding to the angle $\vartheta$. 
In the sub-extremal case, it has been known~\cite{logasymptotics} that 
\begin{equation}
C_1(\vartheta)=-2I^{(1)}[\psi], \ \ \ C_2(\vartheta)=8MI^{(1)}[\psi]
    \label{c1c2rn}
\end{equation}
where $I^{(1)}[\psi]$ is given in terms of the initial data of $\psi$ by the following expression
\begin{equation*}
I^{(1)}[\psi]=\frac{M}{4\pi}\int_{\Sigma_0\cap \mathcal{H}^{+}}\psi+\frac{M}{4\pi}\int_{\Sigma_0}\nabla \psi\cdot n_{\Sigma_0}
\end{equation*}
where the above integrals are taken relative to the respective volume forms. An alternative formula~\cite{yag2020} for $I^{(1)}[\psi]$ allows its computation from future null infinity as follows
\begin{equation}
I^{(1)}[\psi]=\frac{M}{4\pi}\int_{\mathcal{I}^{+}\cap\{\tau\geq 0 \}}\Psi(\tau,\vartheta)d\vartheta d\tau.
    \label{i1nullinfi}
\end{equation}
Note in order to compute $I^{(1)}[\psi]$ via \eqref{i1nullinfi} we need to integrate the radiation field $\Psi(\tau,\vartheta)$ over all angles $\vartheta$. Hence formula \eqref{i1nullinfi} does \textbf{not} provide an anisotropic measurement of $I^{(1)}[\psi]$. It is important however to note that for the full sub-extremal RN family we have
${C_2(\vartheta)}/{C_1(\vartheta)}=-4M$
and this ratio is independent of $\psi$ or $\vartheta$ \cite{logasymptotics}.

Switching now to the extremal RN spacetimes, as mentioned before, previous work \cite{paper4} has shown that 
\begin{equation}
    C_1(\vartheta)=4MH[\psi]-2I^{(1)}[\psi]
    \label{c1ern}
\end{equation}
{\color{black}{\bf The main conjecture that we present numerical evidence for in this work is the following expression for $C_2(\vartheta)$}} in extremal RN:
\begin{equation}
    C_2(\vartheta)=-32M^2H[\psi]+8MI^{(1)}[\psi]
    \label{c2ern}
\end{equation}
for all angles $\vartheta$. {\color{black} Note that assuming $C_2(\vartheta)$ is a linear combination of the two numerical constants $H[\psi]$ and $I^{(1)}[\psi]$ is in part motivated by results derived for $C_2(\vartheta)$ for both sub-extremal RN in~\cite{logasymptotics} as well as sub-leading tails for {\bf Types B/C} ID in~\cite{paper4}}. The relations \eqref{c1ern} and \eqref{c2ern} imply that 
\begin{equation}
H[\psi]=-\frac{1}{16 M^2}\left( C_2(\vartheta)+4MC_1(\vartheta)\right)
\label{hern}
\end{equation}
{\color{black} Thus, our main conjecture motivates the} following expression for a signature of extremality in the context of RN black holes
\begin{equation}
\boxed{s_{\mathcal{I}^{+}}(\vartheta)=-\frac{1}{16 M^2}\left( C_2(\vartheta)+4MC_1(\vartheta)\right)}
\label{s2ern}
\end{equation}
or, equivalently,
\begin{widetext}
\begin{equation}
    s_{\mathcal{I}^{+}}(\vartheta) = -\frac{1}{16 M^2}\cdot \lim_{\tau\rightarrow \infty} \left[ 4M\tau^2\cdot \Psi(\tau,\vartheta) \\
     -  \frac{\tau}{\log(\tau/M)}\cdot\left(\lim_{\tau\rightarrow \infty} \tau^2\cdot \Psi(\tau,\vartheta)\right) 
     + \frac{\tau^3}{\log(\tau/M)}\Psi(\tau,\vartheta) \right]  
\end{equation}
\end{widetext}
and notice that 
\begin{equation*}
    s_{\mathcal{I}^{+}}(\vartheta)= \begin{cases}
0 & \text{for subextremal RN}\\
H[\psi] & \text{for ERN}
    \end{cases}
\end{equation*}
{\bf The signature $s_{\mathcal{I}^{+}}(\vartheta)$ allows us to differentiate extremality from subextremality and in fact to compute (in the extremal case) the horizon hair $H[\psi]$ by knowing the radiation field along a fixed null generator of $\mathcal{I}^{+}$.}

{\color{black} In the next section, we will present numerical evidence that supports this conjecture; in particular, we will numerically verify that $s_{\mathcal{I}^{+}}$ is zero for subextremal black holes (RN and Kerr), and numerically show that it is equal to $H[\psi]$ in the extremal cases.}

\medskip

\textbf{Remark.} \textit{Our analysis relies on the fact that for the full sub-extremal RN we have $\frac{C_2(\vartheta)}{C_1(\vartheta)}=-4M$. Given therefore the expression $C_1(\vartheta)=4MH[\psi]-2I^{(1)}[\psi]$ the worst-case scenario would have been that in extremal RN: $C_2(\vartheta)=-16M^2H[\psi]+8MI^{(1)}[\psi]$. }

\medskip

Finally, it is noteworthy that we can even compute $I^{(1)}[\psi]$ in a similar fashion:
\begin{equation}
I^{(1)}[\psi]=-\left(C_1(\vartheta)+\frac{1}{8M}C_2(\vartheta) \right)
\label{iern}
\end{equation}

\section{Numerical Approach \& Results (RN \& Kerr)}
In this section we summarize our numerical approach and present detailed numerical evidence for the fact that $s_{\mathcal{I}^{+}}(\vartheta)$ equals $H[\psi]$ both the ERN and EK cases, and zero for the sub-extremal cases.

\subsection{Numerical Methodology}
We solve the scalar wave equation for perturbations in ERN and EK black hole backgrounds, focusing on axisymmetric modes ($m=0$). We modify the equation to work in compactified hyperboloidal coordinates $(\tau, \rho, \theta^1, \theta^2)$ that are horizon penetrating and allow for time evolution on hypersurfaces which bring $\ml{I}^+$ to a finite radial coordinate $\rho(\ml{I}^+)=S<\infty$. The relationship between these new coordinates $(\tau,\rho)$ and the spherical Boyer-Lindquist coordinates $(t,r)$ is~\cite{coords}
\begin{eqnarray}
\Omega &=& 1-\frac{\rho}{S}\nn\\
r &=& \frac{\rho}{\Omega(\rho)}\\
v\equiv t+r_*-r &=& \tau+\frac{\rho}{\Omega(\rho)}-\rho-4M\log\Omega(\rho)\nn
\end{eqnarray}
where $S$ denotes the location of $\ml{I}^+$ in hyperboloidal coordinates, $r_*$ is the usual tortoise coordinate and $v$ is the modified advanced time. Note that the angular variables are the same in both coordinate systems.\\

Our numerical implementation scheme entails re-writing the second order PDE in terms of two coupled first-order differential equations. We solve this system using a high-order weighted essentially non-oscillatory (WENO) finite-difference scheme with explicit Shu-Osher time-stepping. In the RN case, we solve a (1+1)D system using the standard decomposition in spherical harmonics, while in the Kerr case we solve (2+1)D system. Details may be found in our previous work~\cite{weno,subir1,subir2,subir3,subir4}. We choose $S=19.0$ and the location of $\ml{H}^+$ such that, $\rho(\ml{H}^+)=0.95$. The initial data is a truncated Gaussian centered at $\rho=(1.0, 1.1, 1.2)$ with a width of $(0.16, 0.22, 0.32)$ and non-zero for $\rho\in[0.95,8]$. This ensures compactly supported initial data but with non-zero support on the $\ml{H}^+$ surface. The angular distribution of this initial data is spherically symmetric i.e. the $\ell=0$ monopole. However, the initial data's angular distribution can be considered as ``generic'' i.e. includes all axisymmetric multipoles, since the higher-multipoles decay rapidly leaving the $\ell=0$ monopole to dominate at late times consistent with Price's Law in both ERN and EK.

Finally, to complete these long duration, high-accuracy and high-precision computations in a reasonable time-frame we make extensive use GPGPU-based parallel computing. For additional details on implementation of such intensive computations on a parallel GPU architecture, we refer the reader to our earlier work on the subject~\cite{weno}. Note that these simulations take significant computational resources to run. Each RN simulations takes a week of wall-clock time, while the Kerr ones take a month. For this reason, we focused our efforts on the extremal cases and did not attempt any near-extremal cases. However, our previous work~\cite{subir2} in that context suggests that for cases in which the black holes are very close to extremality, the extremal case results are a good approximation as a ``transient'' i.e. for early times.

\subsection{Numerical Results}
The numerical solution of the wave equation in compactified coordinates allows us to study the time evolution of the field at $\ml{I}^+$ with ${\theta^{1,2}} = \pi/2$. {\color{black} It is important to note that while we note certain specific values of $\vartheta$ for numerical specificity, our results are actually independent of those values\footnote{This is trivial by construction in ERN, of course, but not in EK.} simply because {\em at late times the $\ell=0$ mode dominates} consistent with Price's law}. To calculate $s_{\mathcal{I}^{+}}(\vartheta)$ for subextremal RN, we first calculate the {\em local power index} (LPI) defined as $LPI\equiv\tau\dot{\psi}/\psi$  at each moment in time for $\tau/M\gtrsim3300$. This is because under the assumptions of the previous sections for the field expansion at $\ml{I}^+$, the LPI is related to the ratio $C_2(\vartheta)/C_1(\vartheta)$ as
\begin{eqnarray}
\tau\left[\frac{\tau\dot{\psi}}{\psi} + 2\right] & = & -\frac{C_2(\vartheta)}{C_1(\vartheta)}\log(\tau/M)  + \frac{C_2(\vartheta)}{C_1(\vartheta)}\nn\\
                                               & + & \mathcal{O}\left(M^2\frac{\log(\tau/M)}{\tau}\right)+\hbox{...}
\end{eqnarray}
Thus, the $LPI$ gives us a direct probe of the ratio $C_2(\vartheta)/C_1(\vartheta)$ by regressing $\tau \times (LPI+2)$ against $\log(\tau/M)$. However, since the $LPI$ requires time derivatives of the field, the regression coefficient (which should theoretically equal $C_2(\vartheta)/C_1(\vartheta)$) as a function of time is noisy. To smooth the time evolution of the estimated ratio $C_2(\vartheta)/C_1(\vartheta)$, we use a smoothing algorithm that employs a simple moving average but with variable window size. Specifically, the algorithm assigns a dynamical window for the moving average which minimizes the variance (a measure of roughness of first order differences of the input time series) while preserving its kurtosis (a measure of tailedness).
\begin{figure}[h]
\centering
\includegraphics[width=0.5\textwidth]{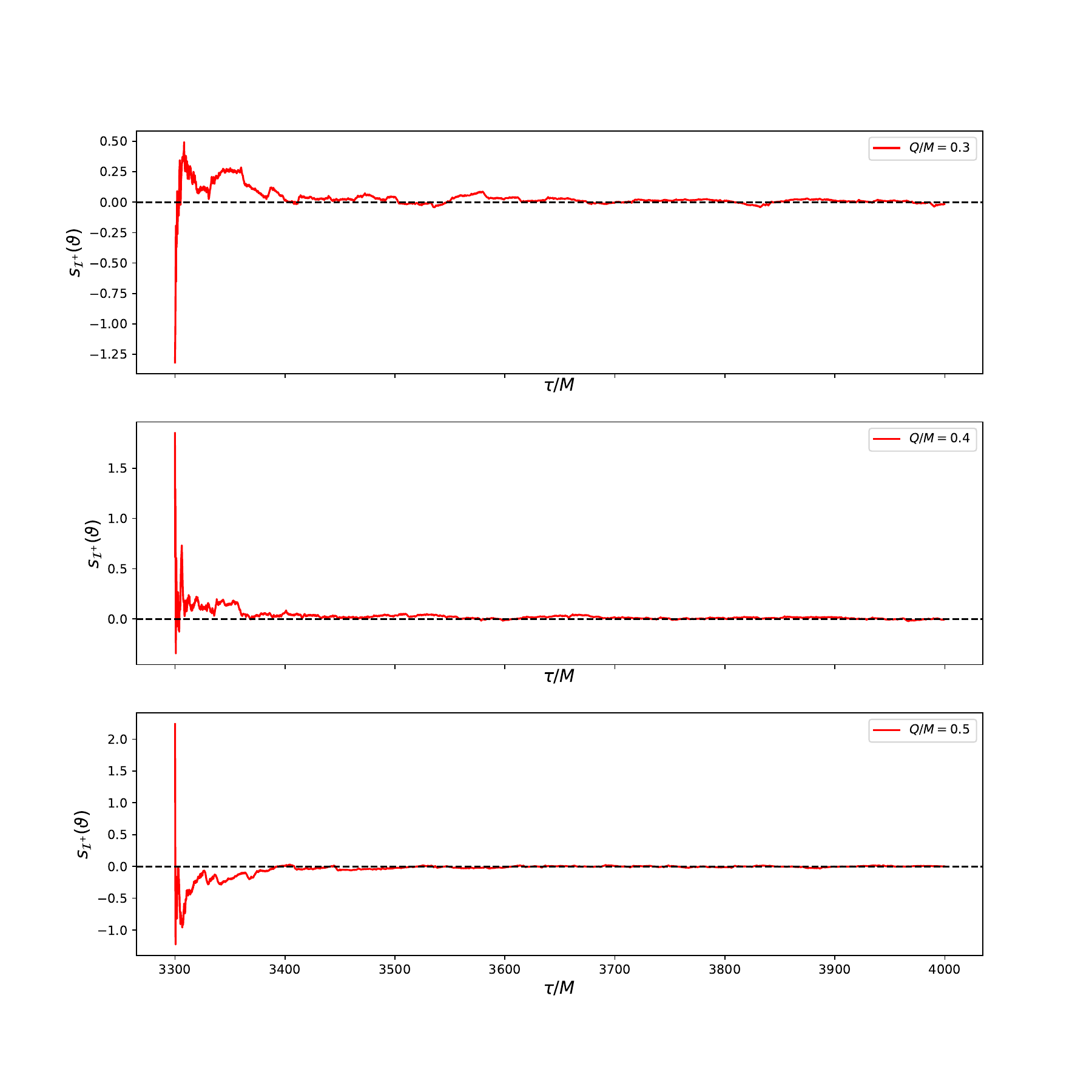}
\caption{\label{fig:subERN}Time evolution of $ s_{\mathcal{I}^{+}}(\vartheta)$ for three sub-extremal RNs, $Q/M=[0.3,0.4,0.5]$ for any value of $\vartheta$. Note that $s_{\mathcal{I}^{+}}(\vartheta)$ asymptotes to $0$ as we evolve the scalar field at $\mathcal{I}^+$.}
\end{figure}
In Fig \ref{fig:subERN}, we plot the estimated values of  $s_{\mathcal{I}^{+}}(\vartheta)$ against time for three different subextremal RN cases, $Q/M=[0.3, 0.4, 0.5]$ at late times ($\tau/M\gtrsim3300$). In each of these cases, it's clear from the numerical data that $s_{\mathcal{I}^{+}}(\vartheta)$ asymptotes to $0$ at late times as desired. This implies that for subextremal BHs, irrespective of their deviation from extremality, $C_2(\vartheta)=-4MC_1(\vartheta)$ as expected~\cite{logasymptotics}.  {\color{black} Recall that in the subextremal cases $C_2(\vartheta)=-4MC_1(\vartheta)$ is known via mathematical derivation~\cite{logasymptotics} and here we are simply offering a numerical illustration of that. These numerical results for the subextremal cases help serve as a test of our numerical approach and thus build a level of confidence in our codes and analysis methodology.}
\begin{figure}[h]
\centering
\includegraphics[width=0.5\textwidth]{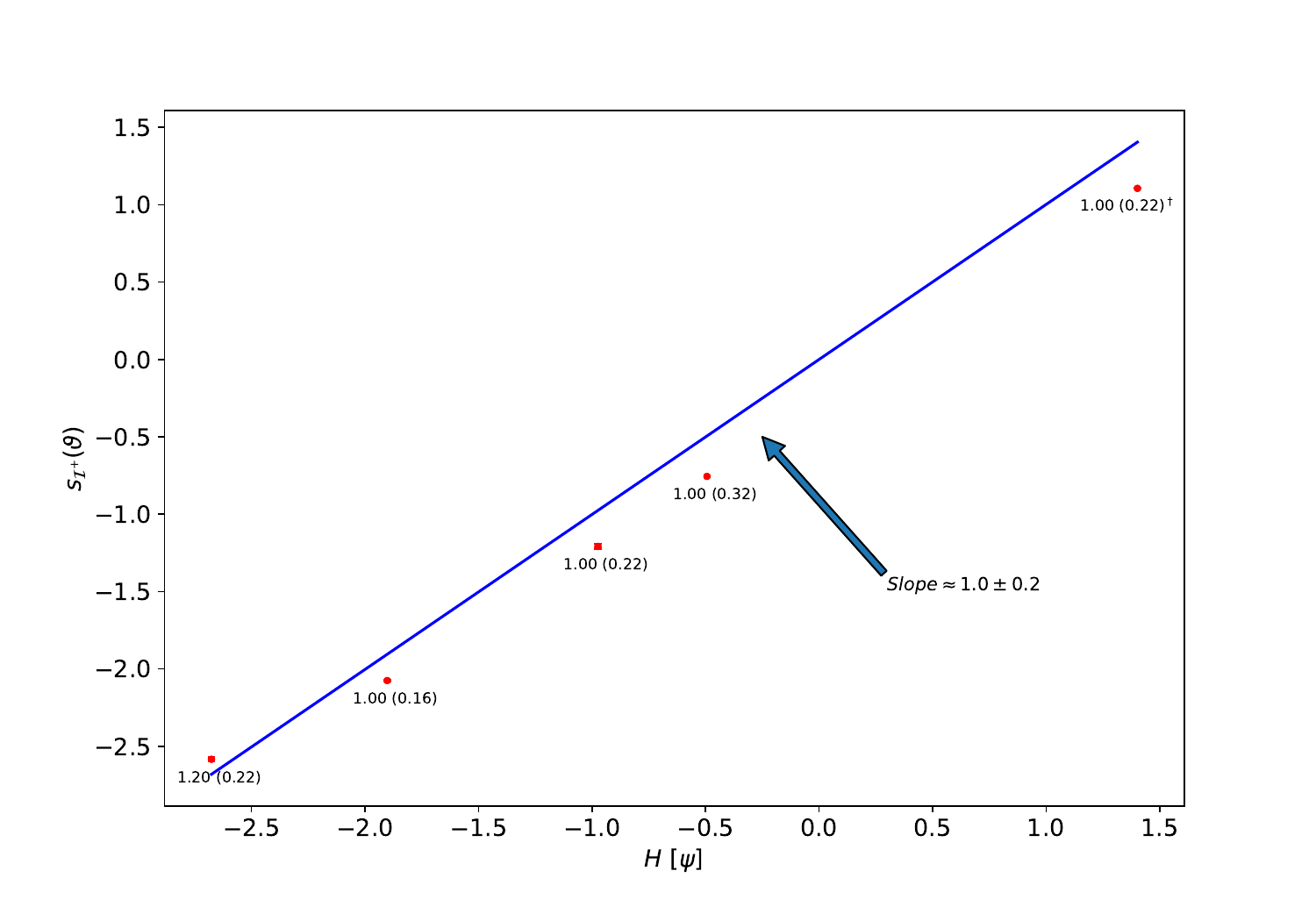}
\caption{\label{fig:ern_s0_s2}Best-fit line demonstrating a linear relationship between the horizon constant $H[\psi]$ and the asymptotically extracted $s_{\mathcal{I}^{+}}(\vartheta)$ for any $\vartheta$ for extremal RN. The slope of the best fit-line is $1.0\pm0.2$. Each data point is labeled by the initial data used for that computation i.e. the location and width of the Gaussian pulse\footnote{The far right data point labeled by the dagger symbol uses initial data in which the field is zero while the momentum is set to the Gaussian pulse.}. The Levenberg–Marquardt fitting errors in each data point are too small to be visible on the scale of the plot.}
\end{figure}
Our proposal is that the quantity $s_{\mathcal{I}^{+}}(\vartheta)$ which depends on the sub-leading terms in the dynamics of the radiation field at $\ml{I}^+$ can indeed provide a signature for extremality. In order to demonstrate this, we determine $s_{\mathcal{I}^{+}}(\vartheta)$ for ERN BHs with different initial conditions and plot it against their respective constant, $H[\psi]$ that depends on initial conditions.

To see the effect of sub-leading terms in the tail behavior for ERN BHs, the simulations were performed up to very late times ($t/M=8000$). Since at each step in the time evolution, we accumulate numerical error in the form of truncation of derivatives, we find that the numerical procedure employed above for sub-extremal BHs didn't work very well for the extremal case -- specifically the $LPI$ was very noisy for the extremal case. Instead, we determined $C_1(\vartheta)$ and $C_2(\vartheta)$ at $\ml{I}^+$ from the time series of the scalar field at $\ml{I}^+$ at late times ($\tau/M\gtrsim5000$) via curve-fitting using the Levenberg–Marquardt algorithm. We plot the relationship between $s_{\mathcal{I}^{+}}(\vartheta)$ and $H[\psi]$ in Fig \ref{fig:ern_s0_s2} that suggests a linear relationship with a slope of the best-fit line being $\approx1$ in accordance with our conjecture. Note that we include the origin $(0,0)$ as a constraint in our linear fitting since $s_{\mathcal{I}^{+}}(\vartheta)$ must go to zero when $H[\psi] = 0$ (see the Appendix).

\begin{figure}[h]
\centering
\includegraphics[width=0.5\textwidth]{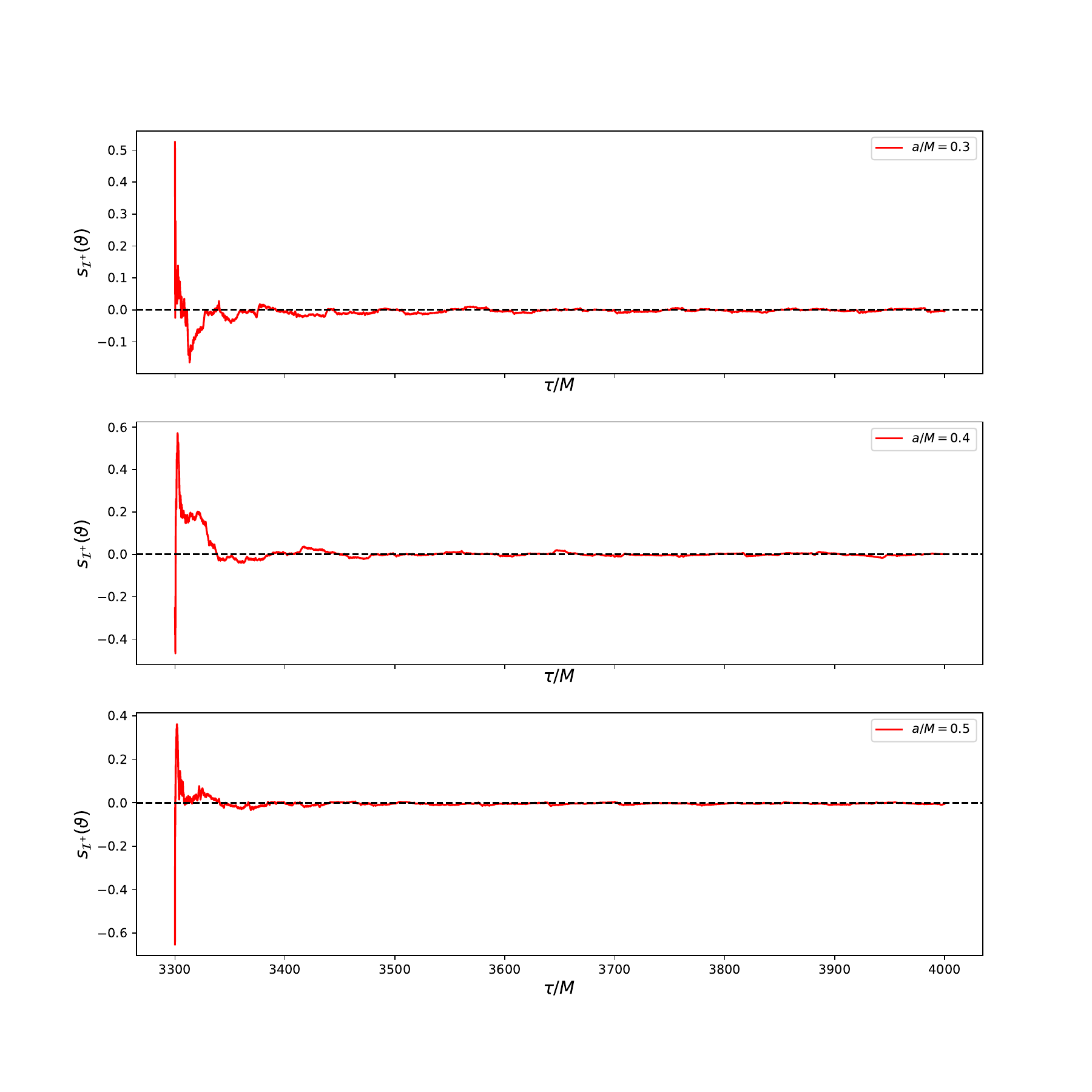}
\caption{\label{fig:subEK} Time evolution of $ s_{\mathcal{I}^{+}}(\vartheta)$ for three sub-extremal Kerr cases, $a/M=[0.3,0.4,0.5]$ for any value of $\vartheta$. Note that $s^{\color{black}Kerr}_{\mathcal{I}^{+}}(\vartheta)$ asymptotes to $0$ as we evolve the scalar field at $\mathcal{I}^+$.}
\end{figure}

\begin{figure}[h]
\centering
\includegraphics[width=0.5\textwidth]{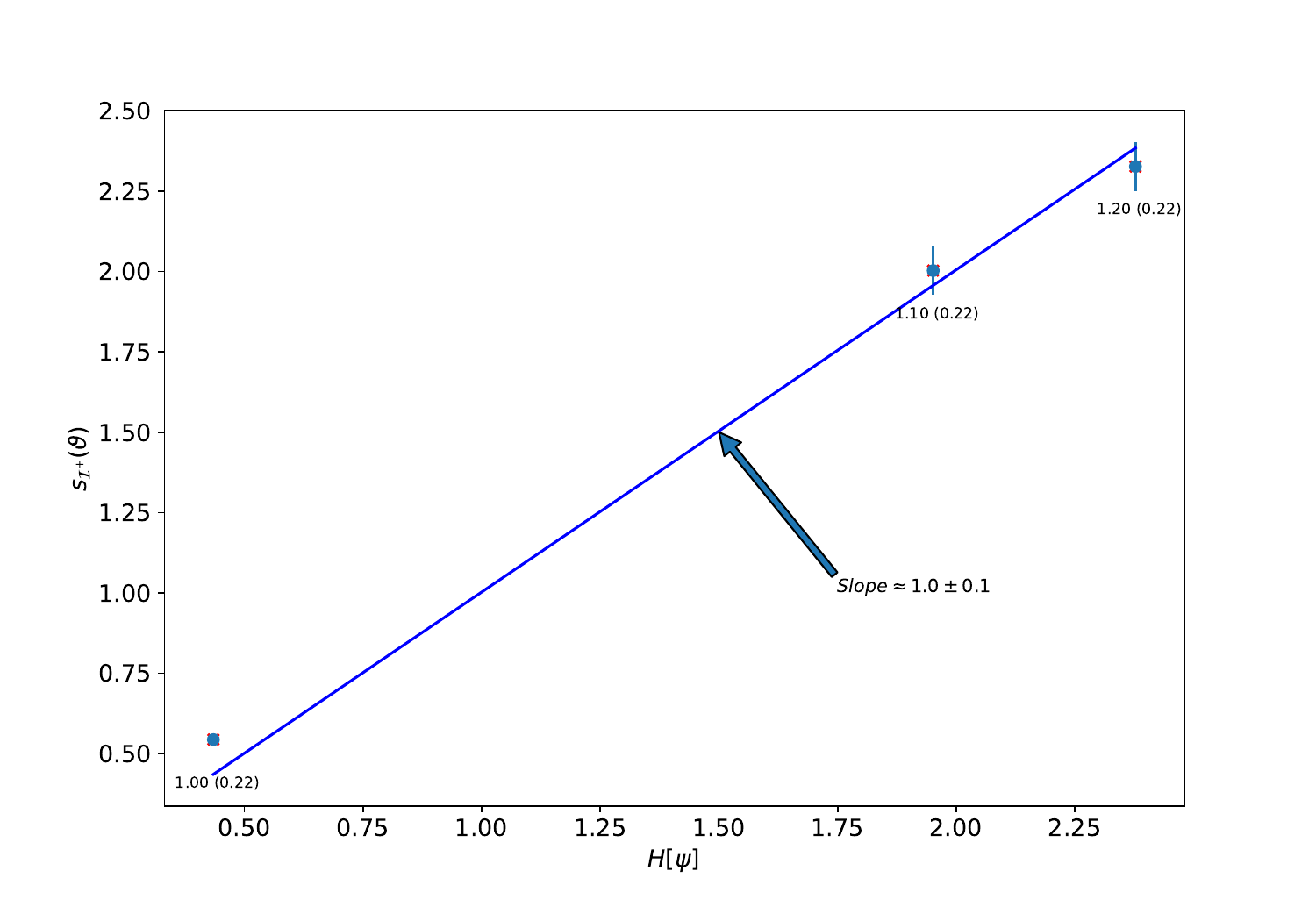}
\caption{\label{fig:ek_s0_s2}Best-fit line demonstrating a linear relationship between the horizon constant $H$ and the asymptotically extracted $s^{\color{black}Kerr}_{\mathcal{I}^{+}}(\vartheta)$ with ${\theta^{1,2}} = \pi/2$ for extremal Kerr black holes. The slope of the best fit-line is $1.0\pm0.1$. Each data point is labelled by the initial data used for that computation i.e. the location and width of the Gaussian pulse. The Levenberg–Marquardt  fitting errors are depicted by the vertical bars on each data point.}
\end{figure}

Finally, we repeated the same analysis for Kerr black holes with axisymmetric scalar perturbations ($m=0$). To start, in Fig \ref{fig:subEK}, we plot the estimated values of  $s_{\mathcal{I}^{+}}(\vartheta)$ against time for three different subextremal Kerr cases, $a/M=[0.3, 0.4, 0.5]$ at late times ($\tau/M\gtrsim3300$). In each of these cases, it's clear from the numerical data that $s_{\mathcal{I}^{+}}(\vartheta)$ asymptotes to $0$ at late times. 
For the extremal Kerr case, we present numerical evidence for a linear relationship between $s_{\mathcal{I}^{+}}$ (redefined below with an extra $1/4$ compared to the ERN case) vs $H[\psi]$ where the quantities are defined similar to RN black holes, i.e. in terms of $C_1(\vartheta)$ and $C_2(\vartheta)$. We  present the results for EK in Fig \ref{fig:ek_s0_s2}. Note that $s_{\mathcal{I}^{+}}$ is redefined for Kerr with an overall $1/4$ compared to (\ref{s2ern}),
\begin{equation}
\boxed{s^{\color{black}Kerr}_{\mathcal{I}^{+}}(\vartheta)=-\frac{1}{64 M^2}\left( C_2(\vartheta)+4MC_1(\vartheta)\right)}
\label{s2k}
\end{equation}
or equivalently,
\begin{widetext}
\begin{equation}
    s^{\color{black}Kerr}_{\mathcal{I}^{+}}(\vartheta) = -\frac{1}{64 M^2}\cdot \lim_{\tau\rightarrow \infty} \left[ 4M\tau^2\cdot \Psi(\tau,\vartheta) \\
     -  \frac{\tau}{\log(\tau/M)}\cdot\left(\lim_{\tau\rightarrow \infty} \tau^2\cdot \Psi(\tau,\vartheta)\right) 
     + \frac{\tau^3}{\log(\tau/M)}\Psi(\tau,\vartheta) \right].  
\end{equation}
\end{widetext} 
This redefinition preserves the property that $s^{\color{black}Kerr}_{\mathcal{I}^{+}}=H[\psi]\neq0$ for EK but $s^{\color{black}Kerr}_{\mathcal{I}^{+}}=0$ for sub-extremal Kerr. 

It is worth noting that the specific numerical values of $s_{\mathcal{I}^{+}}(\vartheta)$ we obtained for both RN and Kerr in the subextremal cases were on the $10^{-2}$ scale at late times. We attribute those values not being closer to zero due to numerical errors arising from our data analysis methodology. This suggests our signature would only be able to conclusively make a claim on extremality if the measured value of $s_{\mathcal{I}^{+}}(\vartheta)$ (and thus $H[\psi]$) is much larger than $10^{-3}$ . However, we do not view this as a serious limitation of our work, since our goal is to offer an ``in principle'' test for extremality and not be constrained by the limitation of our current numerical analysis.

\section{Conclusion (RN \& Kerr)}
\label{conclusion}


In this paper, we have investigated scalar perturbations of the Reissner-Nordstr\"{o}m family and the Kerr family of black holes. Our numerical analysis has allowed us to derive a characteristic excitation for the radiation field at any given fixed angle of future null infinity. The derived expression is non-zero only in the extremal case, where the black hole's charge or angular momentum reaches its maximum value. This observation serves as an important observational signature for extremal black holes, distinguishing them from sub-extremal ones.

Furthermore, we have identified the extremal horizon instability as the source of this excitation in the radiation field. Remarkably, we have demonstrated that the magnitude of the excitation in the radiation field precisely matches the conserved horizon charge of the black hole. This demonstrates that the horizon charge is classically measurable by external observers, including far-away observers, potentially falsifying the no-hair hypothesis in the case of extremal black holes {\color{black} in the context of linearized black hole perturbation theory.}

Our research advances the field of black hole astrophysics by unveiling the observational consequences of extremal black holes and establishing a direct link between the excitation in the radiation field and the conserved horizon charge. These results pave the way for further investigations into the nature and behavior of extremal black holes, opening up new avenues for studying the rich physics associated with these extreme gravitational objects.

\section{Acknowledgements}
\label{ackknow}
The authors wish to thank Som Bishoyi for finding some errors in the formulae presented and double-checking our numerical work. The original manuscript also benefited from the journal peer-review process. G.K. acknowledges support from NSF Grants No.~PHY-2307236 and DMS-2309609. Simulations were performed on the UMass-URI UNITY supercomputer and MIT SuperCloud supported by the Massachusetts Green High Performance Computing Center (MGHPCC). S.A. acknowledge support from the NSERC grant and the Ontario Early Researcher Award.

\appendix
\section{Appendix: Initial data with no support on the horizon}
\begin{figure}[h]
\centering
\includegraphics[width=0.5\textwidth]{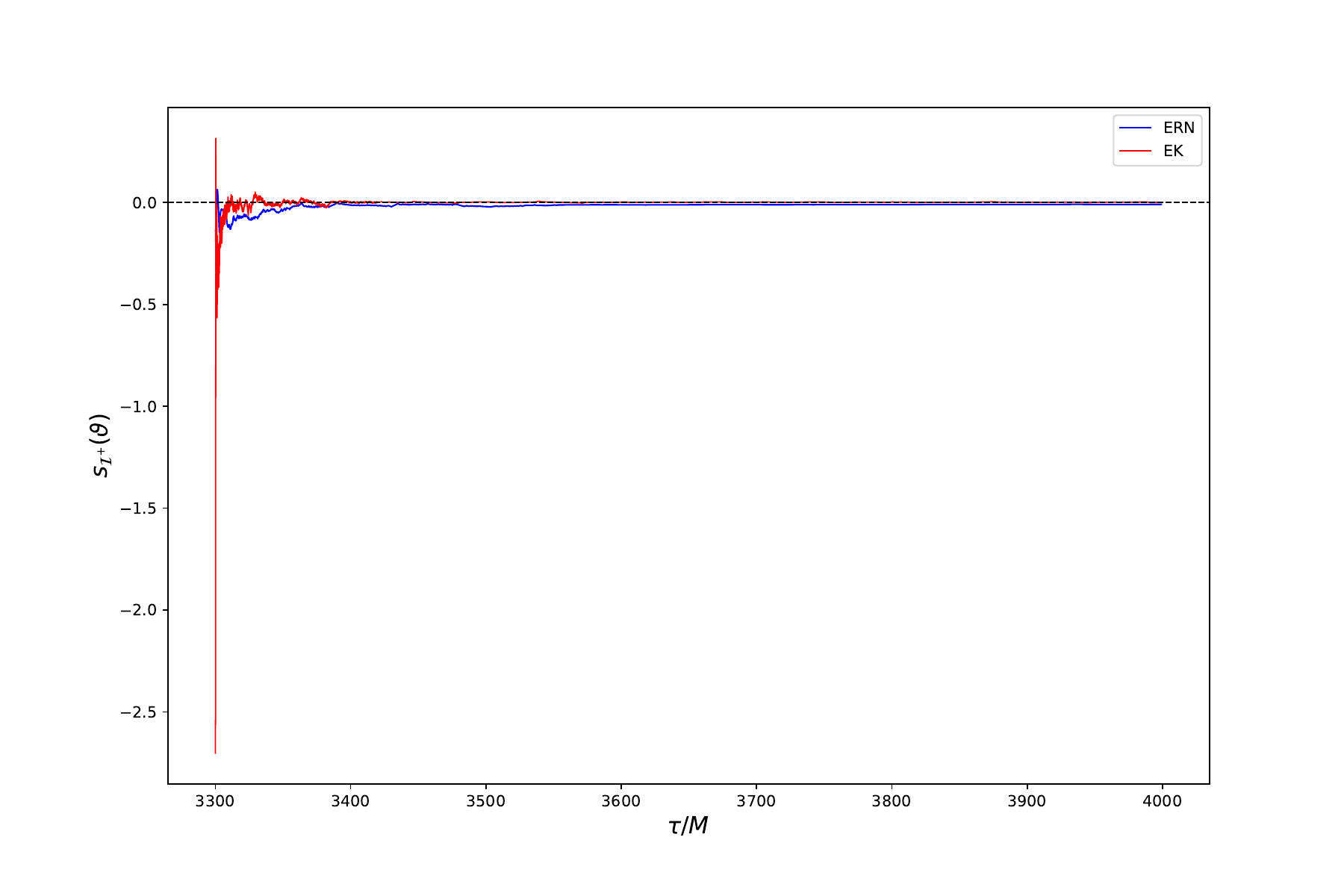}
\caption{\label{fig:no-horizon-id} Time evolution of $s_{\mathcal{I}^{+}}(\vartheta)$ for ERN and EK for initial data with no support on the horizon. Note that $s_{\mathcal{I}^{+}}(\vartheta)$ asymptotes to $0$ as we evolve the scalar field at $\mathcal{I}^+$.}
\end{figure}

In this appendix, we show numerical results that offer numerical confirmation that the proposed signature $s_{\mathcal{I}^{+}}$ works correctly with the initial data lacks support on the horizon for both ERN and EK cases. In Fig.~\ref{fig:no-horizon-id} we show a plot of $s_{\mathcal{I}^{+}}$ for both ERN and EK for an initial Gaussian centered at $\rho = 4.0$ with width $0.22$ and truncated outside the window $[1.0,8.0]$. As mentioned previously, the horizon is located at $\rho = 0.95$ and thus this data has no support on the horizon. It is clear that the signature $s_{\mathcal{I}^{+}}(\vartheta)$ asymptotes to $0$ as we evolve the scalar field. The computational methodology for this Fig.~\ref{fig:no-horizon-id} is the same as that used for Fig.~\ref{fig:subERN}.

\end{document}